\documentclass[
preprint,
nofootinbib,
 amsmath,amssymb,
 aps,
prd,
]{revtex4-1}

\makeatletter
\renewcommand*\l@section{\@dottedtocline{1}{1.5em}{2.3em}}
\renewcommand*\l@subsection{\@dottedtocline{1}{4em}{2em}}
\makeatother

\usepackage{lipsum}
\usepackage{dcolumn}
\usepackage{bm}
\usepackage{color}
\usepackage{graphicx,amsmath,amssymb}
\usepackage{tikz}
\usetikzlibrary{matrix}
\usepackage{slashed}
\usepackage{bbold}
\usepackage{mathrsfs}

\definecolor{awesome}{rgb}{1.0, 0.13, 0.32}
\definecolor{applegreen}{rgb}{0.55, 0.71, 0.0}
\definecolor{darkpastelgreen}{rgb}{0.01, 0.75, 0.24}
\definecolor{azure}{rgb}{0.0, 0.5, 1.0}
\definecolor{fluorescentyellow}{rgb}{0.8, 1.0, 0.0}
\definecolor{guppiegreen}{rgb}{0.1, 0.75, 0.3}
\definecolor{inchworm}{rgb}{0.7, 0.93, 0.36}
\definecolor{richelectricblue}{rgb}{0.03, 0.57, 0.82}
\definecolor{springgreen}{rgb}{0.0, 1.0, 0.5}
\definecolor{mediumcandyapplered}{rgb}{0.89, 0.02, 0.17}
\definecolor{scarlet}{rgb}{1.0, 0.13, 0.0}

\newcommand{\eqnsplit}[1]{\begin{align}\begin{split}#1\end{split}\end{align}}
\newcommand{\eqn}[1]{\begin{align}#1\end{align}}

\newcommand{\lc}{\left(}
\newcommand{\rc}{\right)}
\newcommand{\lcc}{\left[}
\newcommand{\rcc}{\right]}
\newcommand{\lccc}{\left\{ }
\newcommand{\rccc}{\right\} }
\newcommand{\refket}{| \psi_\text{R} \rangle }
\newcommand{\tarket}{| \psi_\text{T} \rangle }

\usepackage{hyperref}
\hypersetup{
    bookmarks=true,         
    unicode=false,          
    pdftoolbar=true,        
    pdfmenubar=true,        
    pdffitwindow=false,     
    pdfstartview={FitH},    
    pdftitle={Title},    
    pdfauthor={I Akal},     
    pdfsubject={Subject},   
    pdfcreator={I Akal},   
    pdfproducer={Producer}, 
    pdfkeywords={keyword1, key2, key3}, 
    pdfnewwindow=true,      
    colorlinks=true,       
    linkcolor=blue,          
    citecolor=guppiegreen,        
    filecolor=blue,      
    urlcolor=blue           
}

\begin{document}
\title{Reflections on Virasoro circuit complexity and Berry phase}

\author{I. Akal}
\affiliation{II. Institute for Theoretical Physics, University of Hamburg,\\
Luruper Chaussee 149, D-22761 Hamburg, Germany}

\email{ibrahim.akal@desy.de}

\date{\today}

\begin{abstract}
Recently, the notion of circuit complexity defined in symmetry group manifolds has been related to geometric actions which generally arise in the coadjoint orbit method in representation theory and play an important role in geometric quantization. On the other hand, it is known that there exists a precise relation between geometric actions and Berry phases defined in group representations.
Motivated by these connections, we elaborate on a relation between circuit complexity and the group theoretic Berry phase. As the simplest setup relevant for holography, we discuss the case of two dimensional conformal field theories. In the large central charge limit, we identify the computational cost function with the Berry connection in the unitary representation of the Virasoro group. 
We then use the latter identification to express the Berry phase in terms of the Virasoro circuit complexity. The former can be seen as the holonomy of the Berry connection along the path in the group manifold which defines the protocol.
In addition, we derive a proportionality relation between Virasoro circuit complexity and the logarithm 
of the inner product between a particularly chosen reference state and the prepared target state. In this sense, the logarithmic formula turns out to be approximating the complexity up to some additive constant if the building blocks of the circuit are taken to be the underlying symmetry gates. Predictions based on this formula have recently been shown to coincide with the holographic complexity proposals and the path integral optimization procedure. The found connections may therefore help to better understand such coincidences. We also discuss that our findings, put together with earlier observations, may suggest a connection between the Virasoro Berry phase and the complexity measure in the path integral optimization proposal.
\end{abstract}

\begin{titlepage}
\maketitle
\thispagestyle{empty}
\end{titlepage}
\tableofcontents

\section{Introduction}
One of the most interesting developments in recent years
have been achieved by coalescing 
ideas from holography, quantum field theory, quantum information and theoretical computer science.
In particular, many of the findings have tremendously helped to better understand the underlying mechanisms behind the anti-de Sitter (AdS)/conformal field theory (CFT) correspondence \cite{Maldacena:1997re,Gubser:1998bc,Witten:1998qj}, which were elusive for so much time since the first proposal of the duality.

As a key insight which has played a pioneering role, it has been shown that there exists a surprising relation between the entanglement properties of the boundary CFT and certain geometric objects in the bulk spacetime \cite{Ryu:2006bv,Ryu:2006ef,Hubeny:2007xt}. 
Such connections are taken as a strong evidence that quantum entanglement plays a fundamental role in the emergence of semiclassical spacetime geometry \cite{VanRaamsdonk:2010pw,Lashkari:2013koa,Faulkner:2013ica,Rangamani:2016dms}.

The found connections turn out to be even more striking when it comes to the physics of black holes (BHs). 
While the entanglement entropy, which is holographically dual to extremal codimension two surfaces in the bulk \cite{Ryu:2006ef,Hubeny:2007xt,Casini:2011kv,Dong:2016hjy}, approaches a constant value with time during BH thermalization, the size of the Einstein-Rosen bridge of the eternal BH in AdS grows further \cite{Hartman:2013qma}.

In view of these findings, it has been argued that the boundary quantity, which evolves even when thermal equilibrium is reached, is quantum computational complexity \cite{Susskind:2014moa}. Motivated by these observations, two different conjectures have been proposed which state that complexity in the boundary quantum field theory (QFT) is related to certain codimension one \cite{Susskind:2014rva,Stanford:2014jda,Roberts:2014isa} and codimension zero \cite{Brown:2015lvg,Brown:2015bva} bulk objects, see also \cite{Couch:2016exn}.
These quantities thus appear to be very suitable for probing the interior regions of BHs in AdS/CFT
and therefore play a substantial role in our efforts to better understand the nature of quantum gravity \cite{Almheiri:2014cka,Papadodimas:2013wnh,Papadodimas:2013jku} which for sure is one of the most important issues in fundamental science. Since their very first proposal, the holographic complexity conjectures not only have advanced our insights into the deep connection between quantum information and the structure of dynamical spacetime, they also conduce as motivation for introducing a notion of complexity in QFTs. 

The concept of computational complexity originates in the field of theoretical computer science. 
For instance, in a quantum circuit model the unitary operation which prepares the target state by mapping an input reference state can be approximated by elementary building blocks in form of quantum gates. 
The complexity is then defined as the minimum number of gates constructing the circuit. In the case when the states are fixed, it is required to optimize over all circuits of the latter sort.

Interesting progress towards defining a notion of complexity in QFT has recently been initiated. 
Many of the ideas are based on the geometric approach proposed by Nielsen and collaborators \cite{nielsen2005geometric,nielsen2006quantum,dowling2008geometry}. According to the latter, determining the optimal circuit is equivalent to finding minimal geodesics in a geometry associated with the space of unitaries which is based on the algebra of the gates. Such techniques thus recast the problem of gate counting, which in general is a highly difficult task, into an optimization problem that can be tackled by using the more conventional technology of differential geometry.

Inspired by this geometric approach, a notion of circuit complexity in QFT has been introduced in \cite{Jefferson:2017sdb} where geodesics are determined in the space of unitaries acting on Gaussian states. 
Many applications and generalizations of this approach have been discussed in recent time \cite{Khan:2018rzm,Hackl:2018ptj,Bhattacharyya:2018bbv,Alves:2018qfv,Guo:2018kzl,Chapman:2018hou,Ali:2018fcz,Ali:2018aon,Jiang:2018nzg,Sinamuli:2019utz,Liu:2019aji,Akal:2019ynl,Bhattacharyya:2019kvj}. 
A slightly different, but related proposal is based on ideas for continuous tensor networks \cite{Haegeman:2011uy,Nozaki:2012zj,Mollabashi:2013lya} and makes use of the Fubini-Study metric \cite{Chapman:2017rqy}. This approach has been utilized in e.g. \cite{Camargo:2018eof,Ali:2018fcz}. 
Further approaches to (circuit) complexity are for instance discussed in \cite{Hashimoto:2017fga,Cotler:2018ufx,Balasubramanian:2018hsu,Ali:2019zcj}. 
It is worth mentioning that the described approaches have  enabled to consult the notion of complexity as a probe for diagnosing quantum chaos, see e.g. \cite{Brown:2017jil,Magan:2018nmu,Ali:2019zcj,Yang:2019iav}.

A different ansatz for defining complexity in QFT has been motivated by the MERA tensor network representation \cite{vidal2008class} of the partition function \cite{Miyaji:2016mxg,Caputa:2017urj,Caputa:2017yrh}. For related ideas also see \cite{Czech:2017ryf}. The approach makes use of the observation, that a discretized Euclidean path integral preparing a given CFT state on flat background should, according to its tensor network interpretation, be optimized by performing the path integral on a Weyl rescaled geometry. This change in the path integral then appears in form of the well known Liouville action which has been proposed as a measure of complexity.

We should note that the complexity associated with the path integral optimization procedure can be reproduced within the framework of the geometric approach to circuit complexity. This has recently been shown in \cite{Camargo:2019isp} by using a path integral formulation developed in the context of tensor networks \cite{Milsted:2018yur,Milsted:2018san}. Note that a similar formulation also appears in \cite{Caputa:2018kdj} which makes explicit use of the symmetries in the CFT. 
Apart from the complexity approaches described above, alternative field theoretic developments have been introduced in \cite{Yang:2017nfn,Belin:2018bpg}.

In the present paper, we focus on the circuit complexity approach in the particular case of two dimensional CFTs.
Instead of considering geometries in the space of unitaries, we follow the proposal in \cite{Magan:2018nmu} which defines the circuit complexity in symmetry group manifolds. 
It has been found that the obtained complexity coincides with the Kirillov geometric action \cite{Caputa:2018kdj}. The latter associated with the coadjoint orbits of the Virasoro group coincides with the Polyakov action of two dimensional gravity \cite{alekseev1989path,alekseev1990geometric}. This interesting connection therefore suggests that optimal quantum computation in two dimensional CFTs may be intimately connected to gravity. 
Geometric actions of the mentioned type appear in the coadjoint orbit method in representation theory \cite{kirillov2004lectures,kostant1970quantization} which also have applications in the context of geometric quantization \cite{witten1988}. 

Apart from the connections above, it is long known  that there exists a relation between geometric actions and Berry phases defined in group representations, cf. e.g. \cite{jordan1988berry,vinet1988invariant,strahov2001berry}. 
As originally shown in \cite{berry1984quantal}, the Berry phase arises due to adiabatic variations of Hamiltonians which induce geometric transformations of Hamiltonian eigenspaces. More precisely, consider a quantum system with some Hamiltonian depending on continuous parameters. In this situation one can define a connection which gives rise to a purely geometric phase. The latter, i.e. the Berry phase, is picked up by quantum states under adiabatic variations of the Hamiltonian parameters. 
Analog studies in QFT have shown that nontrivial Berry phases generally appear in many familiar cases and especially in CFTs with nontrivial conformal manifolds \cite{Baggio:2017aww}. A modified version of the Berry phase has recently been studied in the context of the AdS/CFT as well \cite{Czech:2017zfq}.

In this paper, we tie on the connections described so far, and elaborate on a relation between Virasoro circuit complexity \cite{Caputa:2017urj} and the group theoretic Berry phase defined in the unitary representation of the Virasoro group \cite{Oblak:2017ect}. Our main findings in the present paper are summarized as follows:

$\bullet$ We find that the computational cost function coincides with the (negative) Berry connection in the large central charge limit. This relation we then use to express the Virasoro Berry phase in terms of the circuit complexity plus some constant boundary term which only depends on the endpoints of the path in the group manifold that determines the protocol associated with the circuit.

$\bullet$ We then discuss that the latter relation can be used to derive a proportionality relation between Virasoro circuit complexity and the logarithm of the inner product $\langle \psi_\text{R} | \psi_\text{T} \rangle$ where $| \psi_\text{R} \rangle$ denotes a chosen reference state and $| \psi_\text{T} \rangle$ corresponds to the target state prepared by the Virasoro symmetry circuit. We should note that recent studies of a logarithmic formula of such type, i.e. a modified version of the Fubini-Study distance discussed in \cite{Brown:2017jil}---although motivated from a different point of view \cite{Yang:2019udi}---have led to certain similarities with holographic complexity predictions and the path integral optimization approach. We discuss our results in view of these findings.

$\bullet$ Lastly, we discuss that the identified connections above, together with earlier observations, may suggest an interesting connection between the Berry phase and the complexity measure in the path integral optimization proposal, i.e. the well known Liouville action.

The remaining part of this paper is structured as follows. In \S~\ref{sec:cc}, we discuss basic aspects of circuit complexity and introduce the geometric approach adopted in the present paper. In \S~\ref{sec:vcc}, we review relevant group theoretic properties of the Virasoro group and comment on recent results obtained for computational costs in two dimensional CFTs. In \S~\ref{sec:reps-berry}, we introduce the Berry phase and discuss its generalization to group representations where we particularly focus on the unitary representation of the Virasoro group. Our main findings summarized above are contained in \S~\ref{sec:connects}. The paper is finalized in \S~\ref{sec:disc} with a discussion of the results in view of recent findings.

\section{Circuit complexity}
\label{sec:cc}
\subsection{Quantum circuits}
The notion of computational complexity is a well known concept in theoretical computer science. 
A protocol of mapping a quantum state describing a certain number of qubits to another state is determined by a function which is a unitary operation. 

In a quantum circuit model, the circuit plays the role of the unitary operation and is constructed from quantum gates selected from a fixed set of universal elementary gates. The complexity of the circuit may be defined as the minimal number of gates required to implement the unitary.
Up to some error rate, any unitary operator can be simulated by sequentially applying certain gates picked from the universal elementary gate set. For a given target state, the complexity of any circuit of interest then depends on the initial reference state, a choice of the universal gate set and an error tolerance. 

We may write the (relative) complexity between a given reference state $\refket$ and a target state $\tarket$
as
\eqn{
\mathcal{C}(\refket,\tarket) = \mathrm{min} \lccc \mathcal{C} ( \hat U) : \tarket = \hat U \refket\ \forall\ \hat U \rccc.
}
The complexity of a unitary itself just depends on the choice of the fixed universal set of elementary gates denoted in the following by $\mathcal{G} = \lccc g_1, g_2, \ldots, g_{n-1}, g_n \rccc$.
The complexity of that unitary equals to the minimal number of gates $g_i \in \mathcal{G}$ which implement the quantum circuit, i.e. 
\eqn{
\mathcal{C}(\hat U) = \mathrm{min} \lccc \# \text{gates} \rccc ,\quad \hat U \approx {q_i}_{N} {q_i}_{N-1} \ldots {q_i}_{2} {q_i}_{1}.
\label{eq:C-U}
}
By design,
the gates $g_i$ constructing the circuit are invertible such that $\mathcal{C}(\hat U) = \mathcal{C}(\hat{U}^{-1})$. 

Generally, $\mathcal{G}$ is not unique and one may in principal use different gate sets to construct the circuit that simulates the desired unitary. This explains why any notion of circuit complexity for unitaries mapping between different quantum states necessarily depends on the choice of $\mathcal{G}$. In a quantum computation process which relies on quantum circuits, the choice for the structure of $\mathcal{G}$ defines a substantial part of the problem. 

The concepts discussed so far are well suited for discrete systems. However, for defining a notion of circuit complexity in continuous systems, the idea of counting the number of quantum gates is surely not convenient. An interesting approach for generalizing the concept of complexity to continuous systems via geometrizing the problem has been proposed by Nielsen and collaborators \cite{nielsen2005geometric,nielsen2006quantum,dowling2008geometry}, originally, for systems with a finite dimensional Hilbert space. 

\subsection{Geometrizing complexity}
The geometric approach to complexity for continuous systems basically uses the idea of associating the elementary gate set with some Lie algebra. The unitary operator can be written as a path ordered exponential with an integral argument expressed in terms of a path dependent generator $i \mathcal{H}(s)$ of the algebra which is parametrized by some affine parameter $s$, i.e.
\eqn{
U = \overleftarrow{\mathcal{P}} \exp \lc -  i \int_0^1  ds\ \mathcal{H}(s) \rc. 
\label{eq:hatU-path}
} 
We should note that the path ordering in \eqref{eq:hatU-path} is chosen to be left oriented, i.e. gates at earlier times are applied first, see for instance the string of left oriented gates in \eqref{eq:C-U}. 

The circuit complexity of the unitary, $\mathcal{C}(U)$, can be defined by introducing an inner product for the corresponding algebra. Let $\lccc i \mathcal{O}_I \rccc$ be a basis for the latter, then the inner product may be defined such that it leads to a right invariant Riemannian metric. In terms of the generator from above, which is determined by the following Schr{\"o}dinger equation $\mathcal{H} = i \dot U  U^\dag  = Y^I \mathcal{O}_I$, the inner product yields $G_{I J} Y^I Y^J$. For related ideas also see \cite{Brown:2019whu}. In this framework, the operators $\mathcal{O}_I$ precisely correspond to quantum gates and the control functions\footnote{They specify the velocity vector tangent to the trajectory in the space of unitaries.} $Y^I$ decide whether the gate of type $I$ is switched on or off.
Using this metric, the computational cost of a trajectory $U(\tau)$ equals to its length and is given by
\eqn{
\mathcal{D}(U(\tau)) = \int_0^\tau ds\ \sqrt{ G_{I J} Y^I(s) Y^J(s)}.
\label{eq:D}
}
According to these definitions, the complexity of the unitary can then be expressed as
\eqnsplit{
&\mathcal{C}(U) = \mathrm{min} \lccc \mathcal{D}(U(\tau)) : U(1) \equiv U \wedge U(0) \equiv \mathbb{1}\ \forall\ U(\tau)  \rccc.
}
Thus, the metric choice above associates the usual Riemannian geometry with the length of the trajectory $U$.
However, note that the metric does not need to be Riemannian. More generally, we may write the computational cost \eqref{eq:D} in form of the following functional
\eqn{
\mathcal{D}(U(\tau)) = \int_0^\tau ds\ \mathcal{F}(U(s),V(s))
\label{eq:D-gen}
}
where the cost function $\mathcal{F}(U,V)$ is a local function of $U(s)$ in the space of unitaries $U(s)$ and $V(s)$ is a vector in the tangent space at the point $U(s)$. It is argued that $\mathcal{F}$ must be continuous, positive, homogeneous and should satisfy the triangle inequality, see e.g. \cite{nielsen2005geometric,dowling2008geometry,Guo:2018kzl}. For a recent analysis of different complexity measures from a field theoretic perspective we refer to \cite{Bueno:2019ajd}. Replacing continuity by the criteria of smoothness makes the computational cost \eqref{eq:D-gen} defining a length in Finsler space. The latter corresponds to a class of differential manifolds dressed with a quasi-metric. A Finsler geometry therefore generalizes a Riemannian manifold. In this sense, we are left with the problem of finding the geodesics in the resulting geometry. Once the cost function, or in other words, the inner product is appropriately chosen, the complexity $\mathcal{C}(U)$ is given by the length of the minimal geodesic in the corresponding space of unitaries connecting the identity $\mathbb{1}$ and the target unitary $U$.

Referring to the original situation in which the circuits are constructed by a certain sequence of quantum gates, we may decompose a circuit, expressed in the most generic form as
\eqn{
U(\tau) =  \overleftarrow{\mathcal{P}} \exp \lc -  i \int_0^\tau  ds\ \mathcal{H}(s) \rc,
\label{eq:U-generic}
} 
into infinitesimal gates 
\eqn{
U_\mathcal{H} = e^{-i \mathcal{H}(\tau) d\tau}
\label{eq:inf-gate}
}
where $\mathcal{H}$ is referred to as the instantaneous gate living in the tangent space at $\tau$. Note that the infinitesimal gate \eqref{eq:inf-gate} evolves the initial unitary $U(\tau)$ in \eqref{eq:U-generic} to 
\eqn{
U(\tau + d\tau) = U_\mathcal{H} U(\tau)
\label{eq:ins-gate-eqn}
}
which explains the indications above.
The computation of instantaneous gates, i.e. velocities in the group manifold, is a difficult task and writing a closed form to compute the complexity is not easy to realize. Referring to the recent proposal in \cite{Magan:2018nmu}, we will be discussing in the following that the problem in hand can drastically be simplified by considering submanifolds associated with symmetry groups. 

\subsection{Symmetry gates}
As mentioned before, the instantaneous gate $\mathcal{H}$ can be seen as an element of a Lie algebra associated with a Lie group in some representation. Similarly, one may identify the set of gates as some unitary representation $\mathcal{U}$ of a Lie group $G$ with elements $g$, together with an associated Lie algebra $\mathfrak{g}$. 
In the Hilbert space $\mathscr{H}$, the representation $\mathcal{U}$ would then associate a unitary operator $\mathcal{U}[g]$ for each element of the Lie group $G$. 

Within the framework above, continuous protocols are simply defined by paths $g(\tau)$ in the group manifold $G$ and the instantaneous gate equation \eqref{eq:ins-gate-eqn} can be translated into an analogous group equation of the form \cite{Caputa:2018kdj}
\eqn{
g(\tau + d\tau) = g_{Q_g} \cdot g(\tau),\quad g_{Q_g} \equiv e^{Q_g(\tau) d\tau} 
\label{eq:ins-gate-group}
}
where for two elements $g_1,g_2 \in G$ the product $g_1 \cdot g_2$ corresponds to the group product. The group theoretic infinitesimal gate $g_{Q_g}$ in the homologous equation \eqref{eq:ins-gate-group} depends on the instantaneous gate 
\eqn{
Q_g(s) = \partial_t \big |_{t=s} \lc g(t) \cdot g^{-1}(s)  \rc
\label{eq:Q}
} 
which corresponds to the adjoint transformation of the Maurer-Cartan form. For the Virasoro group the latter will be discussed in more detail in \S~\ref{sec:vcc}. 

Once the above identifications are made, we need to specify the form of the cost function $\mathcal{F}$ in \eqref{eq:D-gen} to compute the computational cost of some path in the group manifold $G$. Motivated by Nielsen's original proposal for finite dimensional quantum spin systems where spin operators act as gates \cite{nielsen2005geometric}, it is natural to consider two different cost functions. These are referred to as the one norm $\mathcal{F}_1$ and the two norm $\mathcal{F}_2$ which may be defined as the first and second moments of the instantaneous gate in the maximally mixed state $\rho_\text{mix} = \mathbb{1}/|\mathscr{H}|$ where $|\mathscr{H}|$ denotes the dimension of the system's Hilbert space. 

As opposed to the former situation, for infinite dimensional systems such as CFTs, or for more general QFTs, the mentioned definitions do not work properly. One may circumvent such obstacles by replacing $\rho_\text{mix}$ in the original complexity metrics by the actual density matrix $\rho(s) = U(s) \rho_0 U^\dag(s)$ for some arbitrary reference state $\rho_0$. Making such a replacement, the two choices take the form  \cite{Caputa:2018kdj}
\eqn{
\mathcal{F}_1[g](s) = |\mathrm{Tr}(\rho_0 \tilde{Q}_g(s))|,\qquad
\mathcal{F}_2[g](s) = \sqrt{- \mathrm{Tr}(\rho_0 \tilde{Q}_g^2(s))},
\label{eq:cost-fcts}
}
where $\tilde Q_g = U^\dag Q_g U$, or 
\eqn{
\tilde Q_g(s) = \partial_t \big |_{t=s} \lc g^{-1}(s) \cdot g(t)  \rc,
\label{eq:Qtilde}
} 
respectively.
The latter is the mentioned Maurer-Cartan form which, together with the instantaneous gate from \eqref{eq:Q}, belongs to the algebra $\mathfrak{g}$ of the Lie group $G$ in some unitary representation $\mathcal{U}$. 
The complexity then is taken to be given by the following computational cost functional
\eqn{
\mathcal{C}_j[g](\tau) \equiv \mathcal{D}_j[g](\tau) = \int_0^\tau ds\ \mathcal{F}_j[g](s)
\label{eq:group-D}
}
where $\mathcal{F}_j$ corresponds to the cost function of interest, see \eqref{eq:cost-fcts}.

\section{Virasoro circuit complexity}
\label{sec:vcc}

\subsection{Virasoro group}
\label{subsec:vir-group}
In order to apply the previous ideas to the Virasoro group, i.e. the centrally extended lifts of orientation preserving $S^1$ diffeomorphisms, we first summarize the relevant group theoretic aspects, before we discuss recent results obtained for computational costs in the Virasoro group manifold.

Let $\text{diff}^+ (S^1)$ be the group of orientation preserving diffeomorphisms on the circle with $S$ being the group of circle rotations.
Any element of $\text{diff}^+ (S^1)$ can be represented by a smooth, periodic map $g$ with $g^\prime (\sigma) > 0 $ which satisfies the constraint $g(\sigma + 2 \pi) = g(\sigma) + 2 \pi\ \forall\ \sigma \in \mathbb{R}$. The identity is taken to be $e(\sigma) = \sigma$ and the inverse $g^{-1}$ is defined such that $g(g^{-1}(\sigma)) = g^{-1}(g(\sigma)) = \sigma$. The group product is given by $g \cdot f \equiv g \circ f$. 

Note that identifying $\sigma$ as some light cone coordinate $x^+$, the group $\text{diff}^+ (S^1)$ of maps $x^+ \mapsto g(x^+)$, more precisely, its universal covering group, becomes the chiral half of the conformal group of a Lorentzian cylinder. In that sense, the covering group, denoted as $\mathrm{diff}( S^1)$ for simplification, is a group of conformal transformations.

The universal central extension of $\mathrm{diff} (S^1)$ then coincides with the Virasoro group. As will be clear below, it is defined as the set
\eqn{
\widehat{\mathrm{diff}}( S^1) \equiv \mathrm{diff}( S^1) \times \mathbb{R}
\label{eq:uce}
}
with paired elements $(g,\alpha)$ where $g \in \mathrm{diff} (S^1)$ and $\alpha \in \mathbb{R}$. The product for the group elements is $(g,\alpha) \cdot (f,\beta) \equiv (g \circ f, \beta + \alpha + \mathsf{C}(g,f))$. The map $\mathsf{C}: \mathrm{diff} (S^1) \times \mathrm{diff} (S^1) \rightarrow \mathbb{R}$ is known as the Bott cocycle which measures the symplectic area of triangles on Virasoro coadjoint orbits \cite{neeb2002central}. 
Using the Bott cocycle identity, the product law implies that the elements have to be inverted according to $(g,\alpha)^{-1} = (g^{-1}, -\alpha) $. 

Note that the Lie algebra of $\mathrm{diff} (S^1)$ is the space of vector fields $X=X(\sigma) \partial_\sigma$ on the circle, i.e. $\mathrm{vect} (S^1)$, with $X(\sigma)$ being a $2\pi$-periodic component. The algebra of the group $\widehat{\mathrm{diff}} (S^1)$ is the extended space $\mathrm{vect} (S^1) \oplus \mathbb{R}$ with paired elements $(X,\alpha)$ where $\alpha \in \mathbb{R}$ and $X$ is as before. 

To show that the Lie algebra of the group $\widehat{\mathrm{diff}} (S^1)$ is the Virasoro algebra, one needs to find the Lie bracket of the group elements. This can be done by evaluating the differential of the adjoint representation of $\widehat{\mathrm{diff}} (S^1)$ defined in  \eqref{eq:uce}.
The adjoint representation of a Lie group $G$ is defined as a map
which associates with $g\in G$ a linear operator
\eqn{
\mathrm{Ad}_g (X) = \frac{d}{dt} \bigg|_{t=0} g \cdot e^{t X} \cdot g^{-1}\quad \forall\ X \in \mathfrak{g}
\label{eq:Ad}
}
that acts on the Lie algebra $\mathfrak{g}$.
The adjoint representation of $\mathrm{diff} (S^1)$ coincides with the transformation law of vector fields on the circle. In the case when the previous group is centrally extended, i.e. $\widehat{\mathrm{diff}} (S^1)$, the equation \eqref{eq:Ad} generalizes to
\eqn{
\mathrm{Ad}_{(g,\alpha)} (X , \beta)
= \lc \mathrm{Ad}_g (X) ,  \beta - \frac{1}{24 \pi}  \int_0^{2 \pi} d\sigma\ Sch[g,\sigma] X(\sigma) \rc.
\label{eq:Ad-centered}
}
As expected, the first entry in \eqref{eq:Ad-centered} corresponds to the centerless case in \eqref{eq:Ad}, whereas the second one is the additional entry due to the presence of the Bott cocycle $\mathsf{C}$ and depends on the usual Schwarzian derivative $Sch[g,\sigma] \equiv g^{\prime\prime\prime}/g^\prime - (\sqrt{3} g^{\prime\prime})^2 / (\sqrt{2} g^\prime)^2$ with $g^\prime \equiv \partial_\sigma g$.

The bracket\footnote{Note that here the Lie bracket of elements of the Virasoro algebra are defined as
$\lcc (X,\alpha),(Y,\beta) \rcc \equiv \frac{d}{ds} \big |_{s=0} \mathrm{Ad}_{(e^{s X},s \alpha)} (Y,\beta)$. } of the Lie algebra can now be obtained by differentiating the adjoint representation \eqref{eq:Ad-centered} and takes the form
\eqn{
\lcc (X,\alpha),(Y,\beta) \rcc
= \lc -[X,Y] , - \frac{1}{2 \pi} \int_0^{2 \pi} d\sigma\ X^{\prime\prime\prime}(\sigma) Y(\sigma) \rc.
\label{eq:Virasoro-bracket}
}
Obviously, it does not receive any contribution from the central terms $\alpha$ and $\beta$.
In addition, let us define the Virasoro generators $\ell_n \equiv (-i e^{i n \sigma} \partial_\sigma, - i \delta_{n,0} / 24)$ and $C \equiv (0,-i)$ where $n \in \mathbb{Z}$. Since the functions $X(\sigma)$ and $Y(\sigma)$ are placed on the circle, these can be Fourier expanded, so that elements of the Virasoro algebra can be written as $(X,\alpha) = \sum_{n \in \mathbb{Z}} X_n \ell_n + i \alpha C$ with the Hermiticity condition for the stress energy tensor, $X_n^* = - X_{-n}$. In a unitary representation $\mathfrak{u}$ of the Virasoro algebra, inserting the latter into the Lie bracket \eqref{eq:Virasoro-bracket} yields the well known relation for the Virasoro generators, i.e.
\eqn{
[L_n,L_m] = (n-m)L_{n+m} + \frac{c}{12} (n^3 - n) \delta_{n,-m}.
\label{eq:Virasoro-algebra}
}
Observe that in \eqref{eq:Virasoro-algebra} the generators are represented by operators $\mathfrak{u}[\ell_n] = L_n$ and $\mathfrak{u}[C]=c \mathbf{1}$ for all $n \in \mathbb{Z}$ where $\mathbf{1}$ is the identity and the definite value $c$ denotes the central charge that commutes with all generators in any irreducible representation, i.e. $[L_n,C] = 0$.

\subsection{Virasoro circuits}
\label{subsec:vir-circuit}
Having discussed the group theoretic properties of the Virasoro group, we 
now turn to computational costs in the symmetry group manifold. Let us consider a unitary quantum circuit built from Virasoro symmetry gates associated with a unitary representation $\mathcal{U}$ of the Virasoro group. Following the preceding discussion, the protocol is defined by a path $g(\tau, \sigma)$ in the group manifold. The initial reference state $\refket$ at $s=0$ is assumed to be associated with $\mathcal{U}[g(0,\sigma)]$, whereas the prepared target state $\tarket$ at $s=\tau$ shall be associated with $\mathcal{U}[g(\tau,\sigma)]$. 

In general, note that the product of the paired group elements of the Virasoro group is an operation which, apart from the composition of functions, includes the addition of central terms together with the Bott cocycle. For the relations discussed here, the additional numerical terms in the group product are taken not to be contributing to the computational cost. This simplification assigns a vanishing complexity for the identity operator. It can be shown that this situation corresponds to the case where the geometric phase in the group manifold vanishes.

To tackle the underlying problem, we first write the circuit in path ordered form
\eqn{
U(\tau) =  \overleftarrow{\mathcal{P}} \exp \lc \int_0^\tau  ds\ Q(s) \rc
\label{eq:U-generic-group}
} 
as in the original integral representation \eqref{eq:U-generic}.
The instantaneous gate $Q$, which belongs to the Virasoro algebra, can be written in terms of the stress tensor $T(\sigma)$ and a velocity function $\epsilon(s,\sigma)$,
\eqn{
Q(s) = \int_0^{2 \pi} \frac{d\sigma}{2 \pi} \epsilon(s,\sigma) T(\sigma).
}
Both quantities $T(\sigma)$ and $\epsilon(s,\sigma)$ can be Fourier expanded on the cylinder in terms of the generators of the Virasoro algebra \eqref{eq:Virasoro-algebra}. Due to symmetry properties of the Virasoro group and the definition of the instantaneous gate according to the group equation \eqref{eq:ins-gate-group}, one can derive an explicit relation between the velocities $\epsilon(s,\sigma)$ and the path $g(s,\sigma)$. This allows to express the cost functions $\mathcal{F}_1$ and $\mathcal{F}_2$ in terms of the path which is defined in the group manifold.

However, in order to determine the explicit form of the cost functions, it is necessary to fix the reference state $\rho_0$ in \eqref{eq:cost-fcts}. Due to natural reasons, one may consider a pure eigenstate created by a primary operator with chiral dimension $h$, i.e.
\eqn{
\rho_0 = | h \rangle \langle h |.
\label{eq:rho0}
}
For the choice from \eqref{eq:rho0} the two cost functions introduced above turn out to be equivalent in the large $c$ limit \cite{Caputa:2018kdj}, i.e. $\mathcal{F}_2  \approx \mathcal{F}_1 \equiv \mathcal{F}$, and take the form
\eqn{
\mathcal{F}(\tau) = \frac{c}{24 \pi} \int_0^{2 \pi} d\sigma\ \frac{\dot g}{g^\prime} \lc \frac{1}{2} - \frac{12 h}{c}  + Sch[g,\sigma]\rc
\label{eq:fin-cost-fct}
}
where $\partial_s g \equiv \dot g$ and $g^\prime$ is defined as before. This is an interesting result which demonstrates that in the present formulation nonuniversal ambiguities in the definition of circuit complexity become erased for holographic CFTs. This will allow us to work with a unique choice for the geometry in the unitary group manifold.
To be noted, recently, an equivalence between two different cost functions has also been achieved by tuning the weighting factor for a certain class of gates in an appropriate way \cite{Akal:2019ynl}. 

Coming back to the present problem, once the explicit form of the cost function \eqref{eq:fin-cost-fct} is known, inserting it into the computational cost functional \eqref{eq:group-D} yields the following expression
\eqn{
\mathcal{C}[g](\tau)
= \frac{c}{24 \pi} \int_0^\tau ds \int_0^{2 \pi} d\sigma\ \frac{\dot g}{g^\prime} \lc \frac{1}{2} - \frac{12 h}{c}  + Sch[g,\sigma]\rc
\label{eq:C-f}
}
for the complexity between the identity operator and symmetry transformations associated with the path $g(\tau,\sigma)$. Recently, a connection to the (geometric) Schwarzian action has also been found for the relative entropy defined in operator algebraic language which can be seen as a measure for the indistinguishability of two states with respect to some reference \cite{Hollands:2019czd}.

Recall that only one copy of the Virasoro group has so far been considered. This is of course not the whole story and one needs to include the contribution from the second copy as well. Following an analogous procedure, the full complexity of the CFT then becomes
\eqn{
\mathcal{C}_\text{CFT} \equiv \mathcal{C}[g] + \mathcal{C}[\bar g]
}
where $\bar g$ corresponds to the path for the left copy of the Virasoro group. 

As already mentioned before, the complexity functional \eqref{eq:C-f} coincides with the Polyakov action of two dimensional gravity. The latter corresponds to the Kirillov geometric action \cite{kirillov2004lectures} on the coadjoint orbits of the Virasoro group. These relations outlined above therefore suggest a connection between optimal quantum computation in two dimensional CFTs and gravity \cite{Caputa:2018kdj}.

\section{Berry phase in group representations}
\label{sec:reps-berry}

\subsection{General aspects}
In this section, we introduce some general aspects of the Berry phase. Afterwards, we turn to its generalization to group representations and, in particular, to the unitary representation of the Virasoro group.

Let $\mathscr{H}$ be the Hilbert space of a given quantum system and $H$ its Hamiltonian which depends on certain external parameters $p$, say the coordinates of some manifold $\mathcal{M}$. In addition, let us assume that $\forall\ p \in \mathcal{M}$ the energies $E_n(p)$, whose spectrum is assumed to be discretized at any $p$, as we indicate by the label $n \in \mathbb{N}$, define a vector bundle with basespace $\mathcal{M}$. Its fibres at each $p$ are subspaces of $\mathscr{H}$ generated by the eigenvectors of the related energy eigenvalues. 

For simplicity, we consider a specific eigenvalue $E_{n}$ which is assumed to be nondegenerate for all $p$. The resulting picture is a bundle with one dimensional fibres. If the state $| \psi_n(p) \rangle$ is a normalized eigenvector of $H(p)$ for each $p$, then for any function $\lambda(p)$ on $\mathcal{M}$, the vectors $e^{i \lambda(p)} | \psi_n(p) \rangle$ are still normalized eigenvectors with the same eigenvalues $E_n(p)$. This is true for infinitely many possible sections of the mentioned complex line bundle which are simply related by $U(1)$ gauge transformations on the basespace. 

Suppose that the initial system is given by $\vert \psi_n (p) \rangle$ and the the external parameters vary in time resulting in a path $\gamma(s)$ on $\mathcal{M}$ where $\gamma(0) = p$. Then, the Hamiltonian depending on $\gamma(s)$ will vary as well. If this change, for instance, is driven adiabatically, means that eigenvalues of $\partial_s H(\gamma)$ are much smaller than $(\Delta E)^2 / \hbar$ with $\Delta E$ being the smallest energy gap on the path $\gamma$, then the 
wave function at time $s$ can be written as \cite{born1928beweis}
\eqn{
\vert \psi(s) \rangle = e^{i \theta_n(s)} \vert \psi_n(\gamma(s)) \rangle.
\label{eq:wave-fct}
}
Or, in other words,
finding the system in the state $\vert \psi_n(\gamma(s)) \rangle$ has probability one.
It can be shown that the phase from above is real and has the form 
\eqn{
\theta_n = \theta_{n,\text{dyn}} +  \theta_{n,\text{geo}}
} 
where $\theta_{n,\text{dyn}}$ is the usual dynamical phase and $\theta_{n,\text{geo}}$ is a purely geometric part. The latter, which is the more interesting contribution, arises due to the fact that $\vert \psi_n \rangle$ depends on $p \in \mathcal{M}$. More precisely, the geometric part can be written as an integral of a Berry connection $A_n$ which can be associated with the state vectors $\vert \psi_n(p) \rangle$ so that
\eqn{
\theta_{n,\text{geo}} = \int_\gamma A_n,
\label{eq:theta-geom}
}
where here $A_n = i \langle \psi_n(\gamma(s)) \vert \partial_s \vert \psi_n(\gamma(s)) \rangle$. In general, the Berry connection can be written in terms of the exterior derivative $d$ of the parameter manifold $\mathcal{M}$, which is $A_n = i \langle \psi_n \vert d \vert \psi_n \rangle$.\footnote{Note that the imaginary unit in \eqref{eq:theta-geom} makes the whole expression a real number. The overlap $\langle \psi_n | \cdot | \psi_n \rangle$ is purely imaginary due to the assumption $\langle \psi_n(p) | \psi_n(p)  \rangle = 1\ \forall\ p \in \mathcal{M}$.} The integral \eqref{eq:theta-geom} is invariant under reparametrization of the path, since $A_n$ is a one form parametrized by some affine parameter. The latter transforms under a local phase change as
\eqn{
A_n \rightarrow A_n - d\alpha
}
where $\alpha(p)$ denotes a function on $\mathcal{M}$. As mentioned, this is nothing but a $U(1)$ gauge field transformation. Hence, generally, the geometric phase turns out to be gauge dependent due to the explicit dependence on the vector phase. However, in case of closed paths $\gamma$, the expression \eqref{eq:theta-geom} becomes a phase independent quantity. This results in a gauge invariant Berry phase of the form
\eqn{
\mathcal{B}_n[\gamma] = \oint_\gamma A_n = \oint_\gamma i \langle \psi_n | d | \psi_n \rangle.
\label{eq:berry-phase1}
}
The latter corresponds to the holonomy of the Berry connection $A_n$ along the path $\gamma$. Alternatively, it can be expressed as the flux of the Berry curvature $F_n = dA_n$ through some two surface with the path $\gamma$ identified as its boundary.
It is the quantity in \eqref{eq:berry-phase1} which corresponds to the phase that gets picked up by the wave function \eqref{eq:wave-fct} between $s = 0$ and $s = \tau$. 

\subsection{Unitary group representations}
\label{subsec:groups}
The notion of a Berry phase can be generalized to unitary group representations. In order to do so, suppose $G$ is a connected Lie group with elements $g,f,$ etc. and its algebra $\mathfrak{g}$ is the tangent space that consists of elements denoted as $X,Y$ and so on. A unitary representation $\mathcal{U}$ of $G$ in some Hilbert space $\mathscr{H}$ associates a unitary operator $\mathcal{U}[g]$ with each element $g$ of the group $G$. Assume a one parameter subgroup generated by an element $X_0 \in \mathfrak{g}$ contained in $G$, then $\mathcal{U}[e^{s X_0}]$ corresponds to an evolution operator with Hamiltonian $H=i\mathfrak{u}[X_0]$ where $\mathfrak{u}$ corresponds to the differential of $\mathcal{U}$ at the identity, i.e. the Lie algebra representation
\eqn{
\mathcal{U}[e^{s X}] = e^{s \mathfrak{u}[X]}\ \forall\ X \in \mathfrak{g}
\label{eq:algebra-def}
}
with $\mathfrak{u}[X]$ being an anti-Hermitian operator for any $X \in \mathfrak{g}$. Generally, the choice of the Hamiltonian is not unique so that any element $g$ may be seen as a change of the reference frame and may associate a family of Hamiltonian operators $\mathcal{U}[g] H \mathcal{U}^{-1} [g]$ with the Lie group $G$ and its unitary representation $\mathcal{U}$. In this way, the group manifold $G$ can be seen as a parameter space with elements $g \in G$ which determine the corresponding Hamiltonian. 

Now, suppose $| \phi \rangle \in \mathscr{H}$ is a normalized eigenvector of $H$ with a nondegenerate, isolated eigenvalue $E$. Then $\forall\ g \in G$ the vector $\mathcal{U}[g] | \phi \rangle$ is an eigenstate of the Hamiltonian
\eqn{
\tilde H \equiv \mathcal{U}[g] H \mathcal{U}^{-1} [g]
\label{eq:tildeH}
}
with the same eigenvalue. Let $g$ be a path in the group manifold $G$, i.e.
\eqn{
g: [0,\tau] \rightarrow G,\ s \mapsto g(s),
}
where $s \in [0,\tau]$. Then the Hamiltonian \eqref{eq:tildeH} evolves in time as well and under the assumption that this evolution is adiabatic, the initial state goes into the final state according to 
\eqn{
\mathcal{U}[g(0)] | \phi \rangle \rightarrow e^{i \theta(\tau)} \mathcal{U} [g(\tau)] | \phi \rangle
}
by picking up a phase 
\eqn{
\theta(\tau) = - \frac{E \tau}{\hbar} + i \int_0^\tau ds\ \langle \phi | \mathcal{U}^\dag [g(s)] \partial_s\ \mathcal{U} [g(s)] | \phi \rangle 
}
which is analogous to \eqref{eq:wave-fct}. As before, the first term is a trivial dynamical part whereas the second contribution is purely geometric due to the dependence on the choice of the path $g(s) \in G$. When the path is closed, i.e. $g(\tau) = g(0)$, this geometric phase results in a Berry phase. By using the unitarity condition, i.e. $\dag \rightarrow -1$, we get
\eqn{
\mathcal{U}^{-1} [g(s)] \partial_s\ \mathcal{U} [g(s)]  = \mathfrak{u} [\Theta_g]
}
with $\mathfrak{u}$ being the Lie algebra representation defined in \eqref{eq:algebra-def}.

The argument $\Theta_g$ is a $\mathfrak{g}$ valued one form on $G$ which is nothing but the mentioned (left) Maurer-Cartan form. The latter associates a vector $\dot g \in T_g G$ with the Lie algebra element and is defined as
 \eqn{
\Theta_g  [\dot g]  (s) \equiv \partial_{t} \big\vert_{t = s} (g^{-1}(s) \cdot g(t))
\label{eq:mcf}
 }
which of course coincides with the definition in \eqref{eq:Qtilde}.
Expressing the Berry connection $A_\phi$ in terms of the Mauer-Cartan form \eqref{eq:mcf} for the respective Lie group, the Berry phase of the group's unitary representation takes the form
\eqn{
\mathcal{B}_\phi[g] = \oint_g A_\phi = \oint_g i \langle \phi | \mathfrak{u}[\Theta_g] | \phi \rangle.
\label{eq:berry-group-closed}
}
Note that the latter relies on the assumption of a closed path $g$ in $G$. 

A generalization of the group theoretic Berry phase \eqref{eq:berry-group-open} can be achieved by demanding 
\eqn{
\mathcal{U} [g(\tau)] | \phi \rangle = e^{i \alpha } \mathcal{U} [g(0)] | \phi \rangle,\quad \alpha \in \mathbb{R}.
\label{eq:berry-group-open-condition}
}
In other words, the states only differ by an element $h$ of the stabilizer
$G_\phi$ which leaves invariant the ray of $|\phi \rangle$, i.e. $g(\tau) = g(0) \cdot h$ where $\mathcal{U}[h] | \phi \rangle = e^{i \alpha} | \phi \rangle \propto | \phi \rangle$. 
The final result takes the form
\eqn{
\mathcal{B}_\phi[g] = \int_f A_\phi - i \log \langle \phi | \mathcal{U} [g^{-1} (0) g(\tau)] | \phi \rangle
\label{eq:berry-group-open}
}
which consists of two pieces. The first (bulk) contribution coincides with \eqref{eq:berry-group-closed}, but it is now valid for an open path $g(s) \in G$. The second (boundary) term---only determined by the path's endpoints---does not depend on the choice of the path $h(s) \in G_\phi$ and cancels the real phase in \eqref{eq:berry-group-open-condition}. Of course, if the path $g(s)$ is closed, then  \eqref{eq:berry-group-open} reduces to the previous Berry phase \eqref{eq:berry-group-closed}. 
However, as long as $g^{-1}(0) g(\tau)$ belongs to $G_\phi$, there is no other assumption required for the path $g(s)$.
Also note that the generalized Berry phase \eqref{eq:berry-group-open} vanishes for any path of the form $g(s) = g(0) \cdot h(s)$ with $h(s) \in G_\phi$.

We see that when varying the representation parameters adiabatically by following a path $g(s)$ in $G$, the Berry phase is not affected by the path itself, but it depends on its projection down to the quotient\footnote{The quotient denoted as $G/G^\prime$ is defined by the equivalence relation $g \sim g \cdot g^\prime$ where  $g \in G$ and  $g^\prime \in G^\prime$.} space $G/G_\phi$. The Berry phase is different from zero if this projection is nontrivial in the sense that it is different from a point. For allowing an open path in the above situation, the projected path in $G/G_\phi$ has to be closed. This criteria elucidates the explicit dependence of the Berry phase on the geometry of the group manifold $G$. 

\section{Connections to circuit complexity}
\label{sec:connects}

\subsection{Virasoro Berry phase}
In this section, which contains the main findings of this paper, we use the foregoing relations to derive an explicit connection between Virasoro circuit complexity and the Berry phase defined in the unitary group representation. Based on this connection, we further show that the complexity in the large central charge limit can be related to the logarithm of the inner product between the initial reference state and the prepared target state.

To begin with, let $\mathfrak{u}$ be a unitary highest weight representation of the Virasoro algebra for which the central charge $c$ is positive and the weight is $h$. The commutator relations for the generators reproduce the algebra in \eqref{eq:Virasoro-algebra}. The highest weight state $\vert h \rangle$ constitutes a basis for the Hilbert space $\mathscr{H}$ which, by definition, satisfies $L_0 | h \rangle = h \vert h \rangle$ and $L_m | h \rangle = 0$ for all $m > 0$, whereas states of the form $L_{-m_1} \ldots L_{-m_k} | h \rangle$ with $m_j > 0$ are known as the descendants. The vacuum state we assume to be $SL(2, \mathbb{R})$ invariant, i.e. $L_{-1} | 0 \rangle = 0$. 

When we take $\mathfrak{u}$ to be the differential of $\widehat{\mathcal{U}}$ at identity, whereas $\widehat{\mathcal{U}}$ represents the Virasoro group $\widehat{\mathrm{diff}}(S^1)$ in \eqref{eq:uce}, then for all paired elements of the algebra, i.e. $\forall\ (X,\alpha) \in \mathrm{vect}(S^1) \oplus \mathbb{R}$, the expression in \eqref{eq:algebra-def} can be generalized to 
\eqn{
\widehat{\mathcal{U}} [(e^{s X} , s \alpha)] = e^{s \mathfrak{u} [(X,\alpha)]}.
\label{eq:algebra-def2}
}
The Virasoro group representation is such that 
\eqn{
\widehat{\mathcal{U}} [(g,\alpha)]= e^{i c \alpha } \mathcal{U}[g],\quad \alpha \in \mathbb{R}
\label{eq:berry-group-open-condition-2}
} 
where $\mathcal{U}$ corresponds to a unitary operator acting on $\mathscr{H}$. It represents the group $\mathrm{diff}(S^1)$ up to a nonzero phase caused due to the central extension. Namely, note that the group product of \eqref{eq:uce} implies that the composition has to be of the form $\mathcal{U}[f] \circ \mathcal{U}[g] = e^{i c \mathsf{C}(f,g) } \mathcal{U}[f \circ g]$ where $\mathsf{C}(f,g)$ again denotes the Bott cocycle.

Following the procedure in \S~\ref{sec:reps-berry}, we can now define a Berry phase associated with symmetry transformations applied to a given energy eigenstate in two dimensional CFT. This can efficiently be conducted by utilizing the relation \eqref{eq:berry-group-open}. 
In order to do so, we need to find out which space the path has to be projected to, so that the formula \eqref{eq:berry-group-open} can be used. Note that for the highest weight vector $| h \rangle$ the stabilizer $G_h$ is the $U(1)$ subgroup of $\mathrm{diff}(S^1)$ generated by $L_0$. General conformal transformations do not act trivially on the state $| h \rangle$.  

Now, if we take $g(s,\cdot) \in \mathrm{diff}(S^1)$, then there is a circle diffeomorphism $\sigma \mapsto g(s,\sigma)\ \forall\ s \in [0,\tau]$. If we demand that the projection of the path $g$ to the quotient $\mathrm{diff}(S^1) / S^1$ is closed, we need to require that $g^{-1}(0) \circ g(\tau)$ defines a rotation operation by an angle $\theta$, i.e. $g^{-1}(0,g(\tau,0)) = \sigma + \theta$, see also \eqref{eq:berry-group-open-condition}.
Then, in close analogy to the formula \eqref{eq:berry-group-open}, the Berry phase with respect to the path $g(s,\cdot)$, which is picked up by the state $\widehat{\mathcal{U}} [(g(s),0)] | h \rangle$, can be defined as
\eqn{
\mathcal{B}_{h,c} [g]
= \int_g A_{h,c} - i \log \lc \langle h |  \widehat{\mathcal{U}}\lcc (g^{-1}(0),0) \cdot (g(\tau),0)  \rcc | h \rangle \rc.
\label{eq:berry-group-open-Vira}
}
Note that $A_{h,c} = i \langle h | \mathfrak{u}[\widehat{\Theta}_g] | h \rangle$ is the Virasoro Berry connection\footnote{To be more precise we should actually write $\mathfrak{u}[\widehat{\Theta}_{g,\alpha}]$. However, we stick to the notation $\mathfrak{u}[\widehat{\Theta}_g]$ and denote the central extension by the hat on top only.} which depends on the centrally extended Maurer-Cartan form $\widehat{\Theta}_{g} [\dot g,0]$ on $\dot g \equiv \dot g(s,\cdot)$ where the dot on top again denotes the derivative with respect to the affine \textit{time} parameter $s$.

The Mauer-Cartan form $\widehat{\Theta}_{g}$ does not receive any contribution from the time independent central term, but relies on the Bott cocycle \cite{Oblak:2017ect}. This is the reason why for the general
path $(g(s), \alpha(s) ) \in \widehat{\mathrm{diff}}(S^1)$ the central piece $\alpha(s)$ has been set zero without loss of generality.
The centrally extended Maurer-Cartan form has a nonzero central piece due to the Bott cocycle and pairs a vector field on the circle and a real number, thus, leading to the Virasoro algebra. In this sense, the Berry phase from \eqref{eq:berry-group-open-Vira} can be seen as the holonomy of its Berry connection on the quotient space with a symplectic Kirillov-Kostant form \cite{kirillov2004lectures,kostant1970quantization} that coincides with the Berry curvature $dA_{h,c}$ \cite{Mickelsson:1987mx}. 

Given some family of diffeomorphisms $g(s,\sigma)$, evaluating the formula \eqref{eq:berry-group-open-Vira} principally yields the Virasoro Berry phase in explicit form. 
This phase is picked up by the primary state undergoing a family of conformal transformations.
In group theoretic language, the Berry phase can be interpreted as a flux of the natural symplectic (Kirillov-Kostant) form on an infinite dimensional coadjoint orbit of the Virasoro group through any surface whose boundary is the continous path in the group manifold \cite{boya2001berry,Oblak:2017ect}.

Nondegenerate symplectic forms of the mentioned kind can be used to define geometric actions \cite{kirillov2004lectures}. The latter, as we previously brought up, generally arise in the coadjoint orbit method in representation theory with applications in geometric quantization \cite{witten1988}. These ideas have recently been discussed in the context of complexity which have led to an interesting connection between Virasoro circuit complexity and the Kirillov geometric action \cite{Caputa:2018kdj}. This could already be seen as an explicit example of a relation between complexity and action as proposed in holography \cite{Brown:2015lvg,Brown:2015bva} and also long before in the quantum computation context \cite{toffoli1998quo}.

Apart from such observations, the connection between geometric actions and Berry phases in group representations has already been pointed out in earlier works. Therefore, it is natural to expect a connection between the latter and the notion of complexity defined in symmetry group manifolds.
In the remaining part of this section, we will be elaborating on these aspects in the particular case of the Virasoro group.

\subsection{Berry connection}
Following the ideas in  \cite{Oblak:2017ect}, it is possible to derive a more explicit expression for the Virasoro Berry phase on basis of the general formula \eqref{eq:berry-group-open-Vira}. It proves useful to start with the computation of the centrally extended Berry connection $A_{h,c}$ which depends on the Maurer-Cartan form $\widehat{\Theta}_{g}$. The computation of $A_{h,c}$ can be achieved by starting with the centreless group $\mathrm{diff}(S^1)$ and then include its central extension based on the Bott cocycle. 
The centrally extended Maurer-Cartan form $\widehat{\Theta}_{g}$ becomes the sum of a centreless term $\Theta_g$, namely, the one contained in \eqref{eq:berry-group-open}, and a second term $\mathsf{D}_g$ due to the central extension which corresponds to the second argument derivative of the Bott cocycle. The final result then reads as
\eqn{
\int_g A_{h,c} = \int_g i \langle h | \mathfrak{u} [ (\Theta_g,0) ] | h \rangle +\int_g  i \langle h | \mathfrak{u} [ (0,\mathsf{D}_g) ] | h \rangle
\label{eq:A-Vir}
}
with the two contributing pieces of the following form \cite{Oblak:2017ect}
\eqn{
\label{eq:A-Vir-piece1}
 \int_g i \langle h | \mathfrak{u} [ (\Theta_g,0) ] | h \rangle 
 &= - \frac{c}{24 \pi} \int_0^\tau ds \int_0^{2 \pi} d\sigma\  \lc  \frac{12 h}{c}  - \frac{1}{2}   \rc \frac{\dot g}{g^\prime},\\
\int_g  i \langle h | \mathfrak{u} [ (0,\mathsf{D}_g) ] | h \rangle 
 &=- \frac{c}{24 \pi} \int_0^\tau ds \int_0^{2 \pi} d\sigma\ \frac{\dot g}{2 g^\prime} \lc \frac{g^{\prime\prime} }{g^\prime} \rc^\prime.
\label{eq:A-Vir-piece2}
}
If we now add together the expressions \eqref{eq:A-Vir-piece1} and \eqref{eq:A-Vir-piece2} and compare with the Virasoro circuit complexity from \eqref{eq:C-f}, we can check that in the large central charge limit the following relation holds
\eqn{
\int_g A_{h,c} =  -\ \mathcal{C}_{h,c}[g](\tau).
\label{eq:A-C}
}
Of course, this limit is particularly interesting, at least from the holographic point of view. So we can use the relation \eqref{eq:A-C} to express the Virasoro Berry phase \eqref{eq:berry-group-open-Vira} in terms of the Virasoro circuit complexity \eqref{eq:C-f} which gives rise to
\eqn{
\mathcal{B}_{h,c}[g](\tau)
=  -\ \mathcal{C}_{h,c}[g](\tau) - i \log \langle h |  \widehat{\mathcal{U}}\lcc (g^{-1}(0),0) \cdot (g(\tau),0)  \rcc | h \rangle.
\label{eq:B-C-piece}
}
Note that the second term on the right hand side of the formula \eqref{eq:B-C-piece} does not depend on the choice of the path, since the only contribution comes from its endpoints. Due to that reason we may consider this boundary piece as an additive constant. Recall that $g$ is generally allowed to be open, see our previous discussion in \S~\ref{sec:reps-berry}. This simply means that the defined Berry phase can be treated as some real number which already hints on a proportionality between the Virasoro circuit complexity \eqref{eq:group-D} and the boundary term being fully determined by the path's endpoints, cf. \eqref{eq:B-C-piece}.

As we have discussed, the Berry phase generally does only exist when the projection of the path to the quotient space is nontrivial. Importantly, the generalized Berry phase \eqref{eq:B-C-piece} not only depends on the extended connection $A_{h,c}$ which lives on the group manifold. Rather, its pullback to the quotient by a section of the principal bundle is needed. It is the second boundary term in \eqref{eq:berry-group-open-Vira} which completes to the full Berry phase in \eqref{eq:B-C-piece}.

\subsection{A simple relation}
Having derived a proportionality relation between the Berry phase in the unitary representation of the Virasoro group and the complexity of the Virasoro symmetry circuit, we can rewrite the expression in \eqref{eq:B-C-piece} even further by working out a more explicit form for the  logarithmic boundary term. Indeed, as we will see, this gives rise to a simple formula which explicitly depends on the overlap between reference state and the target state.

In order to arrive at the latter stage, recall that we previously required a closeness condition for the path $g$ which has led to the Virasoro Berry phase \eqref{eq:berry-group-open-Vira}. Specifically, the assumption was that $g^{-1}(0) \circ g(\tau)$ corresponds to a rotation by an angle $\theta(\tau) = g^{-1}(0,g(\tau,0))$
such that both states differ up to an element of the stabilizer $G_h$.
Recall that the primary reference state has been the highest weight state, i.e. $| \psi_\text{R} \rangle \equiv | h \rangle$, which is the eigenvector of $L_0$. The target state is nothing but the rotated reference state. Formally, we may therefore write 
$| \psi_\text{T} \rangle = U_{\theta(\tau)} | \psi_\text{R} \rangle$ where $U_{\theta(\tau)} \equiv \widehat{\mathcal{U}}[ (g_{\mathrm{rot}(\theta(\tau))} ,0) ] $ with
\eqn{
\widehat{\mathcal{U}}[ (g_{\mathrm{rot}(\theta(\tau))},0) ] | \psi_\text{R} \rangle = \exp\lc i \theta(\tau) \lc h - \frac{c}{24} \rc \rc | \psi_\text{R} \rangle.
\label{eq_R-T}
} 
Using the assumptions above as well as the relation \eqref{eq:berry-group-open-condition-2} which defines the group representation, we can finally bring \eqref{eq:B-C-piece} into the following form
\eqn{
\mathcal{B}_{h,c}[g](\tau)  = - \mathcal{C}_{h,c}[g](\tau) - \vert  \log \langle \psi_\text{R} | \psi_\text{T} \rangle \vert.
\label{eq:B-C-piece-states}
}
The formula \eqref{eq:B-C-piece-states} explicitly relates the Virasoro Berry phase to the circuit complexity plus the norm of the logarithm of the inner product between the highest weight reference state $| \psi_\text{R} \rangle $ and the target state $| \psi_\text{T} \rangle$ prepared by the corresponding Virasoro circuit.
Of course, one may consider a more complicated path different from the one discussed above. 
However, this generally is a highly nontrivial operation.
In the present case, the only property we have required is an open path with a closed projection to the quotient which is fulfilled by the rotation operation by the angle $\theta(\tau)$. 

In addition, let us bring to mind that our discussion so far only focused on one chiral copy of the Virasoro group. As in \S~\ref{subsec:vir-circuit}, for the full CFT circuit complexity $\mathcal{C}_\text{CFT}$ we have to combine left and right sectors.\footnote{Consider a two dimensional CFT on a Lorentzian cylinder. The group of its orientation preserving conformal transformations is a direct product $\mathrm{diff}(S^1) \times \mathrm{diff}(S^1)$. The group's paired elements $(g, \bar g)$ act according to $(x^+ , x^-) \mapsto (g(x^+),\bar g(x^-))$ where $x^\pm$ denote dimensionless light cone coordinates on the cylinder, see \S~\ref{subsec:vir-group}. The maps $g$ and $\bar g$ are independent diffeomorphisms of $\mathbb{R}$ preserving the orientation of the cylinder such that $g^\prime(x^\pm) > 0$.} When the second copy is taken into account, left and right moving Berry phases combine to the full CFT Berry phase
\eqn{
\mathcal{B}_\text{CFT} \equiv \mathcal{B}_{h,c}[g] + \mathcal{B}_{\bar h,\bar c}[\bar g] .
\label{eq:B-CFT}
}
Note that in general the two paths $g$ and $\bar g$ are  not related to each other. 
For instance, one may choose one of the two paths to be the identity. Indeed this is the limit where $\mathcal{B}_\text{CFT}$ reduces to the chiral Virasoro Berry phase \eqref{eq:B-C-piece-states}.
Nevertheless, in a specific case when the left and right central charges do not differ and the logarithmic piece again only depends on the endpoints of the paths, we may write a proportionality relation of the following form
\eqn{
\mathcal{C}_\text{CFT} \propto - 2  | \log  \langle \psi_\text{R} | \psi_\text{T} \rangle  |
\label{eq:rel11}
}
which directly follows from the chiral formula \eqref{eq:B-C-piece-states}.

\section{Discussion}
\label{sec:disc}
The previous derivations have been obtained by using certain simplifications with respect to the unitary transformation. Recall that in the present approach the unitary is associated with a continuous path in the Virasoro group manifold. In general, we may assume a nonzero Berry phase in the unitary group representation as long as the unitary operation corresponds to an open, nontrivial path. Accordingly, we may deduce from the previous formula \eqref{eq:rel11} an equivalent proportionality relation of the form
\eqn{
\mathcal{C}_\text{CFT} \propto - \log |  \langle \psi_\text{R} | \psi_\text{T} \rangle |^2.
\label{eq:rel22}
}
Indeed a state dependent complexity measure of the latter type has recently been discussed in detail, although motivated from a different perspective, cf. \cite{Yang:2019udi}. At first glance, the relation \eqref{eq:rel22} seems to be too simple for being related to a measure of complexity. However, it has been shown, that in certain cases, computations based on a logarithmic formula of the latter kind coincide with predictions relying on the holographic complexity proposals as well as the complexity measure in the path integral optimization proposal for two dimensional CFTs, cf. \cite{Yang:2019udi,Yang:2019iav}. 

These are interesting observations, but at this stage they seem to be a coincidence rather than something profound. In light of such findings, the derivations presented above might indeed shed some light on such coincidences. 
Namely, note that the formula \eqref{eq:rel11} is based on the assumption of a large central charge which of course is the simplest setup being relevant for holography. This might explain why computations based on \eqref{eq:rel22}, which we have extrapolated from \eqref{eq:rel11}, coincide with the holographic expectations as well as the CFT path integral complexity proposal.

Another point is that a complexity proposal such as \eqref{eq:rel22} does not provide any information about 
the gate content. This of course is different from the usual formulation where circuits are composed of elementary gates.
Therefore, the formula \eqref{eq:rel22} itself does not provide any insight into the structure of the corresponding quantum circuit which prepares the target state. However, note that the latter can be related to \eqref{eq:rel11} which, on the contrary, has been developed within the circuit complexity framework. 
Hence, our findings may indicate that a state complexity proposal of the type \eqref{eq:rel22} 
intrinsically encodes a circuit which consists of the Virasoro symmetry gates. 

Closely related, it has been argued that \eqref{eq:rel22} cannot be considered as a complexity measure defined in terms of a local cost function as in \eqref{eq:group-D}, since it only depends on the initial and final states, but not on the path in the group manifold \cite{Bueno:2019ajd} . 
However, let us note that our findings have somewhat demonstrated that, up to an additive constant, one might indeed be able to relate a notion of circuit complexity as defined in \eqref{eq:group-D} to the logarithm of the inner product between the initial reference state and final target state, see \eqref{eq:B-C-piece-states}.

To be noted, the derivations in the present paper are restricted to continuous systems such as CFTs. According to the found connections, the result \eqref{eq:rel11} as well as its equivalent version \eqref{eq:rel22} thus can only be applied to systems of the latter type. 
As discussed recently, the situation changes in the case of discrete qubit systems where the direct application of such complexity measures leads to certain inconsistencies, see e.g. \cite{Yang:2019udi}. In light of these observations, it has been argued that a state dependent proposal of the form \eqref{eq:rel22} cannot be suitable to define a notion of complexity \cite{Bueno:2019ajd}. 

To be more precise, let us consider a discrete system and some protocol which connects an initial $n$-qubit reference state $| \psi_\text{R} \rangle_n$ 
with a final target state $| \psi_\text{T} \rangle_n$. Let the former state be of the form
\eqn{
| \psi_\text{R} \rangle_n \equiv | q_1 \rangle \otimes | q_2 \rangle \otimes \cdots \otimes  | 0 \rangle \otimes | q_n \rangle
} 
where $q_j \in \{ 0,1\}$ for all $j=1,\ldots,n$. Furthermore, suppose the target state is the same as the initial one, but modified by a single qubit flip, e.g.
\eqn{
| \psi_\text{T} \rangle_n \equiv | q_1 \rangle \otimes | q_2 \rangle \otimes  \cdots \otimes | 1 \rangle \otimes | q_n \rangle .
}
Now using the formula \eqref{eq:rel22}, one can verify that the complexity for the protocol connecting the two states would result in an infinite amount of complexity. This surely contradicts any reasonable definition of a notion of complexity, since two almost identical states $| \psi_\text{R} \rangle \approx | \psi_\text{T} \rangle$ (up to one flipped qubit), where the number of qubits $n$ is taken to be sufficiently large, result in a drastic change in complexity. This appears to be the case even though the logarithm introduces some notion of complexity unit. Similar problems arise for the Fubini-Study distance $\mathcal{D}_\text{FS} = \mathrm{arccos} | \langle \psi_\text{R} | \psi_\text{T} \rangle|$ which has been also considered as a measure for the computational cost between two states, see e.g. \cite{Brown:2017jil}.
In order to resolve the problem of infinities it has been proposed that a state complexity of the form \eqref{eq:rel22} may only be applied to continuous systems \cite{Yang:2019udi}. This is exactly what we have shown by deriving the proportionality relation in \eqref{eq:rel11}. The related complexity is associated with a continuous protocol described by the respective path in the Virasoro group manifold. 

As a last point, we want to emphasize that our findings, put together with earlier observations, may suggest an interesting connection between the group theoretic Berry phase and the complexity measure in the path integral optimization proposal, i.e. the well known Liouville action. 
In a similar context, please note that it has recently been discussed whether the Liouville action can be seen as the Euclidean analogue of the Berry phase \cite{Camargo:2019isp}. 

In order to arrive at the mentioned connection above, we start from the formula \eqref{eq:B-C-piece-states}. 
The logarithmic piece we may again rewrite as in \eqref{eq:rel22}. In this way, we can think of a complicated conformal transformation associated with an open path in the group manifold. In addition, the following proportionality relation
\eqn{
| \langle \psi_\text{R}  | \psi_\text{T} \rangle |^2 \propto \exp \lc -S_\text{L} \rc
\label{eq:rel-Log-SL}
}
has recently been derived in 2d CFT \cite{Yang:2019udi}.
Here, $S_\text{L}$ corresponds to the classical on-shell Liouville action which drops from the Euclidean action in the original path integral and the inner
product is defined between the system's ground state and some field operator eigenstate. 
The state dependent part can be subtracted from the complete result such that we still can use the relation \eqref{eq:rel-Log-SL} in our situation where initial and final state may differ from those considered in the original derivation.
If we now bring together \eqref{eq:B-C-piece-states} and \eqref{eq:rel-Log-SL}, what follows is a proportionality relation of the mentioned kind
\eqn{
\mathcal{B}_\text{CFT} \propto    S_\text{L},
}
i.e. between the classical Liouville action and the
Berry phase defined in the unitary representation of the Virasoro group.

\section{Acknowledgements}
I would like to thank Pawel Caputa and Tadashi Takayanagi for useful comments.
I further thank the Yukawa Institute for Theoretical Physics at Kyoto University where parts of this work have been carried out during the workshop Quantum Information and String Theory 2019.

\appendix

\bibliographystyle{JHEP}
\bibliography{article_bib}

\providecommand{\href}[2]{#2}\begingroup\raggedright\begin{thebibliography}{10}

\bibitem{Maldacena:1997re}
J.~M. Maldacena, {\it {The Large N limit of superconformal field theories and
  supergravity}},  {\em Int. J. Theor. Phys.} {\bf 38} (1999) 1113--1133,
  [\href{http://arxiv.org/abs/hep-th/9711200}{{\tt hep-th/9711200}}]. [Adv.
  Theor. Math. Phys.2,231(1998)].

\bibitem{Gubser:1998bc}
S.~S. Gubser, I.~R. Klebanov, and A.~M. Polyakov, {\it {Gauge theory
  correlators from noncritical string theory}},  {\em Phys. Lett.} {\bf B428}
  (1998) 105--114, [\href{http://arxiv.org/abs/hep-th/9802109}{{\tt
  hep-th/9802109}}].

\bibitem{Witten:1998qj}
E.~Witten, {\it {Anti-de Sitter space and holography}},  {\em Adv. Theor. Math.
  Phys.} {\bf 2} (1998) 253--291,
  [\href{http://arxiv.org/abs/hep-th/9802150}{{\tt hep-th/9802150}}].

\bibitem{Ryu:2006bv}
S.~Ryu and T.~Takayanagi, {\it {Holographic derivation of entanglement entropy
  from AdS/CFT}},  {\em Phys. Rev. Lett.} {\bf 96} (2006) 181602,
  [\href{http://arxiv.org/abs/hep-th/0603001}{{\tt hep-th/0603001}}].

\bibitem{Ryu:2006ef}
S.~Ryu and T.~Takayanagi, {\it {Aspects of Holographic Entanglement Entropy}},
  {\em JHEP} {\bf 08} (2006) 045,
  [\href{http://arxiv.org/abs/hep-th/0605073}{{\tt hep-th/0605073}}].

\bibitem{Hubeny:2007xt}
V.~E. Hubeny, M.~Rangamani, and T.~Takayanagi, {\it {A Covariant holographic
  entanglement entropy proposal}},  {\em JHEP} {\bf 07} (2007) 062,
  [\href{http://arxiv.org/abs/0705.0016}{{\tt arXiv:0705.0016}}].

\bibitem{VanRaamsdonk:2010pw}
M.~Van~Raamsdonk, {\it {Building up spacetime with quantum entanglement}},
  {\em Gen. Rel. Grav.} {\bf 42} (2010) 2323--2329,
  [\href{http://arxiv.org/abs/1005.3035}{{\tt arXiv:1005.3035}}]. [Int. J. Mod.
  Phys.D19,2429(2010)].

\bibitem{Lashkari:2013koa}
N.~Lashkari, M.~B. McDermott, and M.~Van~Raamsdonk, {\it {Gravitational
  dynamics from entanglement 'thermodynamics'}},  {\em JHEP} {\bf 04} (2014)
  195, [\href{http://arxiv.org/abs/1308.3716}{{\tt arXiv:1308.3716}}].

\bibitem{Faulkner:2013ica}
T.~Faulkner, M.~Guica, T.~Hartman, R.~C. Myers, and M.~Van~Raamsdonk, {\it
  {Gravitation from Entanglement in Holographic CFTs}},  {\em JHEP} {\bf 03}
  (2014) 051, [\href{http://arxiv.org/abs/1312.7856}{{\tt arXiv:1312.7856}}].

\bibitem{Rangamani:2016dms}
M.~Rangamani and T.~Takayanagi, {\it {Holographic Entanglement Entropy}},  {\em
  Lect. Notes Phys.} {\bf 931} (2017) pp.1--246,
  [\href{http://arxiv.org/abs/1609.01287}{{\tt arXiv:1609.01287}}].

\bibitem{Casini:2011kv}
H.~Casini, M.~Huerta, and R.~C. Myers, {\it {Towards a derivation of
  holographic entanglement entropy}},  {\em JHEP} {\bf 05} (2011) 036,
  [\href{http://arxiv.org/abs/1102.0440}{{\tt arXiv:1102.0440}}].

\bibitem{Dong:2016hjy}
X.~Dong, A.~Lewkowycz, and M.~Rangamani, {\it {Deriving covariant holographic
  entanglement}},  {\em JHEP} {\bf 11} (2016) 028,
  [\href{http://arxiv.org/abs/1607.07506}{{\tt arXiv:1607.07506}}].

\bibitem{Hartman:2013qma}
T.~Hartman and J.~Maldacena, {\it {Time Evolution of Entanglement Entropy from
  Black Hole Interiors}},  {\em JHEP} {\bf 05} (2013) 014,
  [\href{http://arxiv.org/abs/1303.1080}{{\tt arXiv:1303.1080}}].

\bibitem{Susskind:2014moa}
L.~Susskind, {\it {Entanglement is not enough}},  {\em Fortsch. Phys.} {\bf 64}
  (2016) 49--71, [\href{http://arxiv.org/abs/1411.0690}{{\tt
  arXiv:1411.0690}}].

\bibitem{Susskind:2014rva}
L.~Susskind, {\it {Computational Complexity and Black Hole Horizons}},  {\em
  Fortsch. Phys.} {\bf 64} (2016) 44--48,
  [\href{http://arxiv.org/abs/1403.5695}{{\tt arXiv:1403.5695}}]. [Fortsch.
  Phys.64,24(2016)].

\bibitem{Stanford:2014jda}
D.~Stanford and L.~Susskind, {\it {Complexity and Shock Wave Geometries}},
  {\em Phys. Rev.} {\bf D90} (2014), no.~12 126007,
  [\href{http://arxiv.org/abs/1406.2678}{{\tt arXiv:1406.2678}}].

\bibitem{Roberts:2014isa}
D.~A. Roberts, D.~Stanford, and L.~Susskind, {\it {Localized shocks}},  {\em
  JHEP} {\bf 03} (2015) 051, [\href{http://arxiv.org/abs/1409.8180}{{\tt
  arXiv:1409.8180}}].

\bibitem{Brown:2015lvg}
A.~R. Brown, D.~A. Roberts, L.~Susskind, B.~Swingle, and Y.~Zhao, {\it
  {Complexity, action, and black holes}},  {\em Phys. Rev.} {\bf D93} (2016),
  no.~8 086006, [\href{http://arxiv.org/abs/1512.04993}{{\tt
  arXiv:1512.04993}}].

\bibitem{Brown:2015bva}
A.~R. Brown, D.~A. Roberts, L.~Susskind, B.~Swingle, and Y.~Zhao, {\it
  {Holographic Complexity Equals Bulk Action?}},  {\em Phys. Rev. Lett.} {\bf
  116} (2016), no.~19 191301, [\href{http://arxiv.org/abs/1509.07876}{{\tt
  arXiv:1509.07876}}].

\bibitem{Couch:2016exn}
J.~Couch, W.~Fischler, and P.~H. Nguyen, {\it {Noether charge, black hole
  volume, and complexity}},  {\em JHEP} {\bf 03} (2017) 119,
  [\href{http://arxiv.org/abs/1610.02038}{{\tt arXiv:1610.02038}}].

\bibitem{Almheiri:2014cka}
A.~Almheiri and J.~Polchinski, {\it {Models of AdS$_{2}$ backreaction and
  holography}},  {\em JHEP} {\bf 11} (2015) 014,
  [\href{http://arxiv.org/abs/1402.6334}{{\tt arXiv:1402.6334}}].

\bibitem{Papadodimas:2013wnh}
K.~Papadodimas and S.~Raju, {\it {Black Hole Interior in the Holographic
  Correspondence and the Information Paradox}},  {\em Phys. Rev. Lett.} {\bf
  112} (2014), no.~5 051301, [\href{http://arxiv.org/abs/1310.6334}{{\tt
  arXiv:1310.6334}}].

\bibitem{Papadodimas:2013jku}
K.~Papadodimas and S.~Raju, {\it {State-Dependent Bulk-Boundary Maps and Black
  Hole Complementarity}},  {\em Phys. Rev.} {\bf D89} (2014), no.~8 086010,
  [\href{http://arxiv.org/abs/1310.6335}{{\tt arXiv:1310.6335}}].

\bibitem{nielsen2005geometric}
M.~A. Nielsen, {\it A geometric approach to quantum circuit lower bounds},
  {\em arXiv:quant-ph/0502070} (2005).

\bibitem{nielsen2006quantum}
M.~A. Nielsen, M.~R. Dowling, M.~Gu, and A.~C. Doherty, {\it Quantum
  computation as geometry},  {\em Science} {\bf 311} (2006), no.~5764
  1133--1135.

\bibitem{dowling2008geometry}
M.~R. Dowling and M.~A. Nielsen, {\it The geometry of quantum computation},
  {\em Quantum Information \& Computation} {\bf 8} (2008), no.~10 861--899.

\bibitem{Jefferson:2017sdb}
R.~Jefferson and R.~C. Myers, {\it {Circuit complexity in quantum field
  theory}},  {\em JHEP} {\bf 10} (2017) 107,
  [\href{http://arxiv.org/abs/1707.08570}{{\tt arXiv:1707.08570}}].

\bibitem{Khan:2018rzm}
R.~Khan, C.~Krishnan, and S.~Sharma, {\it {Circuit Complexity in Fermionic
  Field Theory}},  {\em Phys. Rev.} {\bf D98} (2018), no.~12 126001,
  [\href{http://arxiv.org/abs/1801.07620}{{\tt arXiv:1801.07620}}].

\bibitem{Hackl:2018ptj}
L.~Hackl and R.~C. Myers, {\it {Circuit complexity for free fermions}},  {\em
  JHEP} {\bf 07} (2018) 139, [\href{http://arxiv.org/abs/1803.10638}{{\tt
  arXiv:1803.10638}}].

\bibitem{Bhattacharyya:2018bbv}
A.~Bhattacharyya, A.~Shekar, and A.~Sinha, {\it {Circuit complexity in
  interacting QFTs and RG flows}},  {\em JHEP} {\bf 10} (2018) 140,
  [\href{http://arxiv.org/abs/1808.03105}{{\tt arXiv:1808.03105}}].

\bibitem{Alves:2018qfv}
D.~W.~F. Alves and G.~Camilo, {\it {Evolution of complexity following a quantum
  quench in free field theory}},  {\em JHEP} {\bf 06} (2018) 029,
  [\href{http://arxiv.org/abs/1804.00107}{{\tt arXiv:1804.00107}}].

\bibitem{Guo:2018kzl}
M.~Guo, J.~Hernandez, R.~C. Myers, and S.-M. Ruan, {\it {Circuit Complexity for
  Coherent States}},  {\em JHEP} {\bf 10} (2018) 011,
  [\href{http://arxiv.org/abs/1807.07677}{{\tt arXiv:1807.07677}}].

\bibitem{Chapman:2018hou}
S.~Chapman, J.~Eisert, L.~Hackl, M.~P. Heller, R.~Jefferson, H.~Marrochio, and
  R.~C. Myers, {\it {Complexity and entanglement for thermofield double
  states}},  \href{http://arxiv.org/abs/1810.05151}{{\tt arXiv:1810.05151}}.

\bibitem{Ali:2018fcz}
T.~Ali, A.~Bhattacharyya, S.~Shajidul~Haque, E.~H. Kim, and N.~Moynihan, {\it
  {Time Evolution of Complexity: A Critique of Three Methods}},
  \href{http://arxiv.org/abs/1810.02734}{{\tt arXiv:1810.02734}}.

\bibitem{Ali:2018aon}
T.~Ali, A.~Bhattacharyya, S.~Shajidul~Haque, E.~H. Kim, and N.~Moynihan, {\it
  {Post-Quench Evolution of Distance and Uncertainty in a Topological System:
  Complexity, Entanglement and Revivals}},
  \href{http://arxiv.org/abs/1811.05985}{{\tt arXiv:1811.05985}}.

\bibitem{Jiang:2018nzg}
J.~Jiang and X.~Liu, {\it {Circuit Complexity for Fermionic Thermofield Double
  states}},  {\em Phys. Rev.} {\bf D99} (2019), no.~2 026011,
  [\href{http://arxiv.org/abs/1812.00193}{{\tt arXiv:1812.00193}}].

\bibitem{Sinamuli:2019utz}
M.~Sinamuli and R.~B. Mann, {\it {Holographic Complexity and Charged Scalar
  Fields}},  \href{http://arxiv.org/abs/1902.01912}{{\tt arXiv:1902.01912}}.

\bibitem{Liu:2019aji}
F.~Liu, R.~Lundgren, P.~Titum, J.~R. Garrison, and A.~V. Gorshkov, {\it
  {Circuit Complexity across a Topological Phase Transition}},
  \href{http://arxiv.org/abs/1902.10720}{{\tt arXiv:1902.10720}}.

\bibitem{Akal:2019ynl}
I.~Akal, {\it {Weighting gates in circuit complexity and holography}},
  \href{http://arxiv.org/abs/1903.06156}{{\tt arXiv:1903.06156}}.

\bibitem{Bhattacharyya:2019kvj}
A.~Bhattacharyya, P.~Nandy, and A.~Sinha, {\it {Renormalized Circuit
  Complexity}},  \href{http://arxiv.org/abs/1907.08223}{{\tt
  arXiv:1907.08223}}.

\bibitem{Haegeman:2011uy}
J.~Haegeman, T.~J. Osborne, H.~Verschelde, and F.~Verstraete, {\it
  {Entanglement Renormalization for Quantum Fields in Real Space}},  {\em Phys.
  Rev. Lett.} {\bf 110} (2013), no.~10 100402,
  [\href{http://arxiv.org/abs/1102.5524}{{\tt arXiv:1102.5524}}].

\bibitem{Nozaki:2012zj}
M.~Nozaki, S.~Ryu, and T.~Takayanagi, {\it {Holographic Geometry of
  Entanglement Renormalization in Quantum Field Theories}},  {\em JHEP} {\bf
  10} (2012) 193, [\href{http://arxiv.org/abs/1208.3469}{{\tt
  arXiv:1208.3469}}].

\bibitem{Mollabashi:2013lya}
A.~Mollabashi, M.~Nozaki, S.~Ryu, and T.~Takayanagi, {\it {Holographic Geometry
  of cMERA for Quantum Quenches and Finite Temperature}},  {\em JHEP} {\bf 03}
  (2014) 098, [\href{http://arxiv.org/abs/1311.6095}{{\tt arXiv:1311.6095}}].

\bibitem{Chapman:2017rqy}
S.~Chapman, M.~P. Heller, H.~Marrochio, and F.~Pastawski, {\it {Toward a
  Definition of Complexity for Quantum Field Theory States}},  {\em Phys. Rev.
  Lett.} {\bf 120} (2018), no.~12 121602,
  [\href{http://arxiv.org/abs/1707.08582}{{\tt arXiv:1707.08582}}].

\bibitem{Camargo:2018eof}
H.~A. Camargo, P.~Caputa, D.~Das, M.~P. Heller, and R.~Jefferson, {\it
  {Complexity as a novel probe of quantum quenches: universal scalings and
  purifications}},  {\em Phys. Rev. Lett.} {\bf 122} (2019), no.~8 081601,
  [\href{http://arxiv.org/abs/1807.07075}{{\tt arXiv:1807.07075}}].

\bibitem{Hashimoto:2017fga}
K.~Hashimoto, N.~Iizuka, and S.~Sugishita, {\it {Time evolution of complexity
  in Abelian gauge theories}},  {\em Phys. Rev.} {\bf D96} (2017), no.~12
  126001, [\href{http://arxiv.org/abs/1707.03840}{{\tt arXiv:1707.03840}}].

\bibitem{Cotler:2018ufx}
J.~Cotler, M.~R. Mohammadi~Mozaffar, A.~Mollabashi, and A.~Naseh, {\it
  {Renormalization Group Circuits for Weakly Interacting Continuum Field
  Theories}},  \href{http://arxiv.org/abs/1806.02831}{{\tt arXiv:1806.02831}}.

\bibitem{Balasubramanian:2018hsu}
V.~Balasubramanian, M.~DeCross, A.~Kar, and O.~Parrikar, {\it {Binding
  Complexity and Multiparty Entanglement}},  {\em JHEP} {\bf 02} (2019) 069,
  [\href{http://arxiv.org/abs/1811.04085}{{\tt arXiv:1811.04085}}].

\bibitem{Ali:2019zcj}
T.~Ali, A.~Bhattacharyya, S.~S. Haque, E.~H. Kim, N.~Moynihan, and J.~Murugan,
  {\it {Chaos and Complexity in Quantum Mechanics}},
  \href{http://arxiv.org/abs/1905.13534}{{\tt arXiv:1905.13534}}.

\bibitem{Brown:2017jil}
A.~R. Brown and L.~Susskind, {\it {Second law of quantum complexity}},  {\em
  Phys. Rev.} {\bf D97} (2018), no.~8 086015,
  [\href{http://arxiv.org/abs/1701.01107}{{\tt arXiv:1701.01107}}].

\bibitem{Magan:2018nmu}
J.~M. Magan, {\it {Black holes, complexity and quantum chaos}},  {\em JHEP}
  {\bf 09} (2018) 043, [\href{http://arxiv.org/abs/1805.05839}{{\tt
  arXiv:1805.05839}}].

\bibitem{Yang:2019iav}
R.-Q. Yang and K.-Y. Kim, {\it {Time evolution of the complexity in chaotic
  systems: concrete examples}},  \href{http://arxiv.org/abs/1906.02052}{{\tt
  arXiv:1906.02052}}.

\bibitem{vidal2008class}
G.~Vidal, {\it Class of quantum many-body states that can be efficiently
  simulated},  {\em Physical review letters} {\bf 101} (2008), no.~11 110501.

\bibitem{Miyaji:2016mxg}
M.~Miyaji, T.~Takayanagi, and K.~Watanabe, {\it {From path integrals to tensor
  networks for the AdS/CFT correspondence}},  {\em Phys. Rev.} {\bf D95}
  (2017), no.~6 066004, [\href{http://arxiv.org/abs/1609.04645}{{\tt
  arXiv:1609.04645}}].

\bibitem{Caputa:2017urj}
P.~Caputa, N.~Kundu, M.~Miyaji, T.~Takayanagi, and K.~Watanabe, {\it {Anti-de
  Sitter Space from Optimization of Path Integrals in Conformal Field
  Theories}},  {\em Phys. Rev. Lett.} {\bf 119} (2017), no.~7 071602,
  [\href{http://arxiv.org/abs/1703.00456}{{\tt arXiv:1703.00456}}].

\bibitem{Caputa:2017yrh}
P.~Caputa, N.~Kundu, M.~Miyaji, T.~Takayanagi, and K.~Watanabe, {\it {Liouville
  Action as Path-Integral Complexity: From Continuous Tensor Networks to
  AdS/CFT}},  {\em JHEP} {\bf 11} (2017) 097,
  [\href{http://arxiv.org/abs/1706.07056}{{\tt arXiv:1706.07056}}].

\bibitem{Czech:2017ryf}
B.~Czech, {\it {Einstein Equations from Varying Complexity}},  {\em Phys. Rev.
  Lett.} {\bf 120} (2018), no.~3 031601,
  [\href{http://arxiv.org/abs/1706.00965}{{\tt arXiv:1706.00965}}].

\bibitem{Camargo:2019isp}
H.~A. Camargo, M.~P. Heller, R.~Jefferson, and J.~Knaute, {\it {Path integral
  optimization as circuit complexity}},  {\em Phys. Rev. Lett.} {\bf 123}
  (2019), no.~1 011601, [\href{http://arxiv.org/abs/1904.02713}{{\tt
  arXiv:1904.02713}}].

\bibitem{Milsted:2018yur}
A.~Milsted and G.~Vidal, {\it {Tensor networks as path integral geometry}},
  \href{http://arxiv.org/abs/1807.02501}{{\tt arXiv:1807.02501}}.

\bibitem{Milsted:2018san}
A.~Milsted and G.~Vidal, {\it {Geometric interpretation of the multi-scale
  entanglement renormalization ansatz}},
  \href{http://arxiv.org/abs/1812.00529}{{\tt arXiv:1812.00529}}.

\bibitem{Caputa:2018kdj}
P.~Caputa and J.~M. Magan, {\it {Quantum Computation as Gravity}},  {\em Phys.
  Rev. Lett.} {\bf 122} (2019), no.~23 231302,
  [\href{http://arxiv.org/abs/1807.04422}{{\tt arXiv:1807.04422}}].

\bibitem{Yang:2017nfn}
R.-Q. Yang, {\it {Complexity for quantum field theory states and applications
  to thermofield double states}},  {\em Phys. Rev.} {\bf D97} (2018), no.~6
  066004, [\href{http://arxiv.org/abs/1709.00921}{{\tt arXiv:1709.00921}}].

\bibitem{Belin:2018bpg}
A.~Belin, A.~Lewkowycz, and G.~Sárosi, {\it {Complexity and the bulk volume, a
  new York time story}},  {\em JHEP} {\bf 03} (2019) 044,
  [\href{http://arxiv.org/abs/1811.03097}{{\tt arXiv:1811.03097}}].

\bibitem{alekseev1989path}
A.~Alekseev and S.~Shatashvili, {\it Path integral quantization of the
  coadjoint orbits of the virasoro group and 2-d gravity},  {\em Nuclear
  Physics B} {\bf 323} (1989), no.~3 719--733.

\bibitem{alekseev1990geometric}
A.~Alekseev and S.~Shatashvili, {\it From geometric quantization to conformal
  field theory},  {\em Communications in mathematical physics} {\bf 128}
  (1990), no.~1 197--212.

\bibitem{kirillov2004lectures}
A.~A. Kirillov, {\em Lectures on the orbit method}, vol.~64.
\newblock American Mathematical Soc., 2004.

\bibitem{kostant1970quantization}
B.~Kostant, {\it Quantization and unitary representations},  in {\em Lectures
  in modern analysis and applications III}, pp.~87--208.
\newblock Springer, 1970.

\bibitem{witten1988}
E.~Witten, {\it Coadjoint orbits of the virasoro group},  {\em Comm. Math.
  Phys.} {\bf 114} (1988), no.~1 1--53.

\bibitem{jordan1988berry}
T.~F. Jordan, {\it Berry phases and unitary transformations},  {\em Journal of
  mathematical physics} {\bf 29} (1988), no.~9 2042--2052.

\bibitem{vinet1988invariant}
L.~Vinet, {\it Invariant berry connections},  {\em Physical Review D} {\bf 37}
  (1988), no.~8 2369.

\bibitem{strahov2001berry}
E.~Strahov, {\it Berry’s phase for compact lie groups},  {\em Journal of
  Mathematical Physics} {\bf 42} (2001), no.~5 2008--2022.

\bibitem{berry1984quantal}
M.~V. Berry, {\it Quantal phase factors accompanying adiabatic changes},  {\em
  Proceedings of the Royal Society of London. A. Mathematical and Physical
  Sciences} {\bf 392} (1984), no.~1802 45--57.

\bibitem{Baggio:2017aww}
M.~Baggio, V.~Niarchos, and K.~Papadodimas, {\it {Aspects of Berry phase in
  QFT}},  {\em JHEP} {\bf 04} (2017) 062,
  [\href{http://arxiv.org/abs/1701.05587}{{\tt arXiv:1701.05587}}].

\bibitem{Czech:2017zfq}
B.~Czech, L.~Lamprou, S.~Mccandlish, and J.~Sully, {\it {Modular Berry
  Connection for Entangled Subregions in AdS/CFT}},  {\em Phys. Rev. Lett.}
  {\bf 120} (2018), no.~9 091601, [\href{http://arxiv.org/abs/1712.07123}{{\tt
  arXiv:1712.07123}}].

\bibitem{Oblak:2017ect}
B.~Oblak, {\it {Berry Phases on Virasoro Orbits}},  {\em JHEP} {\bf 10} (2017)
  114, [\href{http://arxiv.org/abs/1703.06142}{{\tt arXiv:1703.06142}}].

\bibitem{Yang:2019udi}
R.-Q. Yang, Y.-S. An, C.~Niu, C.-Y. Zhang, and K.-Y. Kim, {\it {To be
  unitary-invariant or not?: a simple but non-trivial proposal for the
  complexity between states in quantum mechanics/field theory}},
  \href{http://arxiv.org/abs/1906.02063}{{\tt arXiv:1906.02063}}.

\bibitem{Brown:2019whu}
A.~R. Brown and L.~Susskind, {\it {The Complexity Geometry of a Single Qubit}},
   \href{http://arxiv.org/abs/1903.12621}{{\tt arXiv:1903.12621}}.

\bibitem{Bueno:2019ajd}
P.~Bueno, J.~M. Magan, and C.~S. Shahbazi, {\it {Complexity measures in QFT and
  constrained geometric actions}},  \href{http://arxiv.org/abs/1908.03577}{{\tt
  arXiv:1908.03577}}.

\bibitem{neeb2002central}
K.-H. Neeb, {\it Central extensions of infinite-dimensional lie groups},  in
  {\em Annales de l'institut Fourier}, vol.~52, pp.~1365--1442, 2002.

\bibitem{Hollands:2019czd}
S.~Hollands, {\it {Relative entropy for coherent states in chiral CFT}},
  \href{http://arxiv.org/abs/1903.07508}{{\tt arXiv:1903.07508}}.

\bibitem{born1928beweis}
M.~Born and V.~Fock, {\it Beweis des adiabatensatzes},  {\em Zeitschrift
  f{\"u}r Physik} {\bf 51} (1928), no.~3-4 165--180.

\bibitem{Mickelsson:1987mx}
J.~Mickelsson, {\it {String Quantization on Group Manifolds and the Holomorphic
  Geometry of Diff S1 / S1}},  {\em Commun. Math. Phys.} {\bf 112} (1987) 653.

\bibitem{boya2001berry}
L.~J. Boya, A.~M. Perelomov, and M.~Santander, {\it Berry phase in homogeneous
  k{\"a}hler manifolds with linear hamiltonians},  {\em Journal of Mathematical
  Physics} {\bf 42} (2001), no.~11 5130--5142.

\bibitem{toffoli1998quo}
T.~Toffoli, {\it Quo vadimus? --- much hard work is still needed},  {\em
  Physica D: Nonlinear Phenomena} {\bf 120} (1998), no.~1-2 1--11.

\end{thebibliography}\endgroup

\end{document}